\documentclass[preprint,nobibnotes,preprintnumbers,superscriptaddress,showpacs,sort&compress,longbibliography]{revtex4-1} 

\usepackage{graphicx}		
\usepackage{titlesec}
\usepackage{blindtext}
\usepackage{xfrac}		
\usepackage{subcaption}
\usepackage{multirow}	
\usepackage{gensymb}
\usepackage{amssymb}
\usepackage{balance}
\usepackage{soul}
\usepackage[normalem]{ulem}	
\usepackage{smartref}	
\usepackage{nomencl}
\usepackage[T1]{fontenc}
\makenomenclature

\usepackage[colorlinks=true,linkcolor=blue,citecolor=red,urlcolor=blue]{hyperref}		
																								
\newcommand{\comment}[1]{} 
\makeatletter
\renewcommand{\p@subsection}{}
\renewcommand{\p@subsubsection}{}
\makeatother

\usepackage{setspace}

\usepackage{etoolbox}
\renewcommand\nomgroup[1]{%
  \item[\bfseries
  \ifstrequal{#1}{A}{ }{%
  \ifstrequal{#1}{G}{Greek symbols}{%
  \ifstrequal{#1}{B}{Abbreviations }{}}}%
]}

\begin{document}																													 
%
																												

\newcommand{\kvec}{\mbox{{\scriptsize {\bf k}}}}
\newcommand{\lvec}{\mbox{{\scriptsize {\bf l}}}}
\newcommand{\qvec}{\mbox{{\scriptsize {\bf q}}}}

\newcommand*{\PN}[1]{{\color[rgb]{0, 0, 1}{#1}}}
\newcommand*{\WE}[1]{{\color[rgb]{1, 0, 0}{#1}}}
\newcommand*{\AT}[1]{{\color[rgb]{0, 0.5, 0.5}{#1}}}
\newcommand*{\PK}[1]{{\color[rgb]{0.949, 0.149, 0.835}{#1}}}

\def\eq#1{(\ref{#1})}
\def\sec#1{\hspace{1mm}section \ref{#1}}
\def\chap#1{\hspace{1mm}section \ref{#1}}
\def\fig#1{\hspace{1mm}Fig. \ref{#1}}
\def\figur#1{\hspace{1mm}Figure \ref{#1}}
\def\tab#1{\hspace{1mm}Table \ref{#1}}

\title{Turbulent separation control with a tilted wavy wall: a promising approach for energy savings in aerodynamic systems}

\author{Piotr Kamiński}
\affiliation{Department of Thermal Machinery, Czestochowa University of Technology, Al. Armii Krajowej 21, 42-200 Czestochowa, Poland}

\author{Artur Tyliszczak}
\affiliation{Department of Thermal Machinery, Czestochowa University of Technology, Al. Armii Krajowej 21, 42-200 Czestochowa, Poland}

\author{Witold Elsner}
\affiliation{Department of Thermal Machinery, Czestochowa University of Technology, Al. Armii Krajowej 21, 42-200 Czestochowa, Poland}

\author{Pawe{\l} Niegodajew} \email{pawel.niegodajew@pcz.pl}
\affiliation{Department of Thermal Machinery, Czestochowa University of Technology, Al. Armii Krajowej 21, 42-200 Czestochowa, Poland}

\date{July 9, 2024} 
\begin{abstract}
\footnotesize{Recently, Kami{\'n}ski et al.~\cite{kaminski2024numerical} demonstrated that a two-dimensional streamwise waviness with carefully selected amplitude and period can be effectively used in postponement of a flow separation at high Reynolds number which is out of reach for other commonly known passive flow control strategies. This paper demonstrates that this approach can be substantially improved by introducing a novel type of surface waviness characterized by tilting the subsequent waves. The research is performed by applying the Large Eddy Simulation (LES) method allowing for a detailed analysis of both instantaneous and time-averaged flow characteristics. It is shown that the tilting eliminates the occurrence of separation zones in waviness troughs, which minimizes the shape drag and maximizes the wall shear stress. In particular, compared to an optimal classical sinusoidal waviness shape the drag coefficient drops 14\%. Simultaneously, 70\% better performance in separation control is achieved. Moreover, it is also shown that when this issue is a priority, even further improvement can be achieved, though at the expense of greater drag.
}
\\



%

\end{abstract}
\maketitle
%
%
%

\clearpage
\section{Introduction} \label{intro}

The energy efficiency of aerodynamic systems, such as an airplane wing, strongly depends on its near-wall flow dynamics. When a flow experiences an adverse pressure gradient (APG), which is present on the suction side of an airfoil, the separation of the turbulent boundary layer (TBL) is likely to occur when the streamwise component of the wall-shear stress $\tau_w = \mu (dU/dy)$ (where $\mu$ is the dynamic viscosity and $dU/dy$ is the near-wall velocity gradient) reaches zero locally. Detachment of TBL from a surface is accompanied with a significant increase in total drag, resulting in a substantial energy loss and consequently, increased fuel consumption~\cite{lin1996low, tanarro2020effect}. Other examples where separation can significantly affect energy efficiency include the flow inside a diffuser \cite{azad1996turbulent, salehi2017computation, yadegari2020numerical}, around a wind turbine blade \cite{devinant2002experimental, cui2020simulations, abdolahifar2022comprehensive, xu2017delayed, wang2018influence} or within axial or centrifugal pumps used as turbines~\cite{shojaeefard2024analyzing,nguyen2024co}. Ricco et al.~\cite{RICCO2021100713}, in their review paper, suggested that considerable energy savings, particularly in aerodynamic systems, may be obtained by introducing appropriate flow control methods, which can contribute up to $15\%$ of total drag reduction. Bearing this in mind, the control of separation is critical to minimize energy losses and thus reduce fuel consumption. This scenario has motivated scientists to search for new or to improve already known flow control techniques.

In general, there are two types of flow control approaches: active and passive~\cite{ashill2005review, ghaemi2020passive, moghaddam2017active}. Active control involves systems, often quite complex, that require an external energy source~\cite{collis2004issues}. Due to the necessity for external input and high cost, these methods are often less appealing from the perspective of energy savings. On the other hand, passive control strategies, which typically involve surface geometry modifications or introducing roughness, do not require external energy input and are generally more cost-effective, which makes them significantly more attractive for practical applications. A number of passive approaches have so far been developed and the following ones are the most commonly employed in practice: dimples~\cite{tay2018drag, tay2015mechanics, gattere2022dimples, aoki2012mechanism, tay2019drag, azlan2023passive}; vortex generators~\cite{koike2004research, aider2010drag}, which are employed in various types of systems, i.e., airfoil surfaces, helicopter rotors~\cite{gibertini2015helicopter}, wind turbine blades~\cite{zhu2019dynamic, wang2017flow}, and car bodies~\cite{islam2013drag}. Also a considerable attention is paid to investigating some more unconventional control methods, such as modifying the airfoil surface shape by introducing slots~\cite{belamadi2016aerodynamic, coder2020design}, slats~\cite{zaki2022effects, wang2019effects}, Gurney flaps~\cite{ye2023numerical, ni2021impacts}, or placing microcylinders near a blade leading edge~\cite{WANG2018101, mostafa2022quantitative, wang2023wake, zhong2023dynamic}. Quite often, nature itself suggests solutions that are optimal for the given flow scenario and inspires researchers and engineers to employ them in practice. Good examples of such solutions are wingtip devices known as winglets, which are inspired by bird wings~\cite{guerrero2012biomimetic, wu2018experimental, guerrero2020variable, zhang2023comparative, jeong2022efficient, khaled2019investigation}; riblets~\cite{lowrey1991preliminary} or adaptive flaps~\cite{hao2020performance} reflecting the shapes of bird feathers~\cite{chen2013biomimetic, yossri2023evaluation}; and various types of microstructures borrowed from the skin of aquatic animals~\cite{schlieter2016mechanical, lang2017separation, wu2018experimental}. 

Unfortunately, as suggested in Refs.~\cite{lissaman1983low, xie2022investigation}, the effectiveness of the outlined passive approaches is limited to flows experiencing laminar or transitional conditions. For high Reynolds numbers, when the flow becomes fully turbulent, employment of, in particular, an artificial roughness is not recommended as it may lead to increased total drag and even to earlier separation~\cite{mcmasters1979low, Wu_Piomelli_2018}. In light of these facts, it appears that the use of active methods is the most reasonable choice to prevent TBL separation at high $Re$.

In the experimental study by Dróżdż et al.~\cite{drozdz2021effective} devoted to air flow exposed to APG conditions, the authors showed that by precisely adjusting the amplitude $A$ and the period length $\lambda$ of a two-dimensional (2D) streamwise waviness on a wall surface, it is possible to significantly increase $\tau_{w}$, thereby delaying TBL separation. They suggested that the amplitude of the waviness should be scaled to the size of the inner layer, meaning $A$ should increase proportionally to the TBL thickness $\delta$ found, in the same area, for a configuration with a flat wall. Simultaneously, the corrugation crests should not penetrate the outer region of TBL. Notably, the increase in $\tau_w$, which ensures a delay of turbulent separation, was achieved in a fully turbulent flow regime, where most passive flow control methods are ineffective. Worth noting is that the streamwise waviness has found so far some other applications such as reducing the leading-edge interaction noise of an airfoil~\cite{casalino2019aeroacoustic, teruna2022numerical} or enhancement of heat transfer process~\cite{bhardwaj2015influence, abbaspour2024numerical}.

Recently, Kami{\'n}ski et al.~\cite{kaminski2024numerical} performed Large Eddy Simulation (LES) studies of a flow over 2D streamwise wavy wall (WW) under APG and ZPG conditions at the friction Reynolds number equal to $Re_{\tau}=\frac{u_{\tau}\delta}{\nu}=2500$ (where $u_{\tau}$ is the friction velocity, $\delta$ is the boundary layer thickness and $\nu$ is the kinematic viscosity). In agreement with the experimental research of Dróżdż et al.~\cite{drozdz2021effective} it has been proven that by using a WW shape one can increase $\tau_{w}$ locally, thereby postponing the point of TBL separation. Interestingly, enhancement of $\tau_{w}$ was reported only under APG conditions. The authors also demonstrated that the benefit of augmented $\tau_w$ comes at the cost of increased total drag in the area occupied by the waviness. 

The studies of Kami{\'n}ski et al.~\cite{kaminski2024numerical} and Dróżdż et al.~\cite{drozdz2021effective}, though demonstrated a large potential of WW in the control of $\tau_w$, were limited only to a sinusoidal type of the waviness in which the adjustable parameter was the amplitude $A$ of the sine wave. In the present study, we maintain a constant $A/\delta$ ratio as in \cite{kaminski2024numerical,drozdz2021effective} and additionally modify the waviness by introducing streamwise and counter-streamwise tilting to its sinusoidal shape.  
The research is performed by applying the large eddy simulation (LES) method. It is shown that by employing a particular type of tilted waviness, one can achieve not only a significant $\tau_w$ enhancement compared to sinusoidal waviness, but also a considerable reduction in the local drag coefficient $C_d=2F_d/(\rho U_{e,in}^2 S_{ref})$ (where $S_{ref}$ is the reference area, $\rho,\, U_{e,in}$ denote the density and edge velocity taken as 99\% of the free-stream velocity, and $F_d$ stands for the drag force computed as the surface integral of pressure and the wall-shear stress). Both of these findings are highly attractive from the perspective of energy efficiency in aerodynamic systems and serve as valuable clues indicating areas for further development of efficient passive separation control.

The manuscript is organised as follows. The description of the experimental configuration constituting the research object adopted for LES is presented in Section~\ref{setup}. Details of the boundary conditions and numerical methods are given in Section~\ref{methods}. The results obtained are presented in Section~\ref{results} and include a qualitative description of the flow field followed by comprehensive analysis of the time-averaged solutions aimed at finding the sources of reported performance improvement of the new wall shape. The conclusions and suggestions for future work are provided in Section \ref{conclusions}.

\newpage
\section{Research object \label{setup}}

\begin{figure}[h!] 
    \centering
    \includegraphics[width=0.95\linewidth]{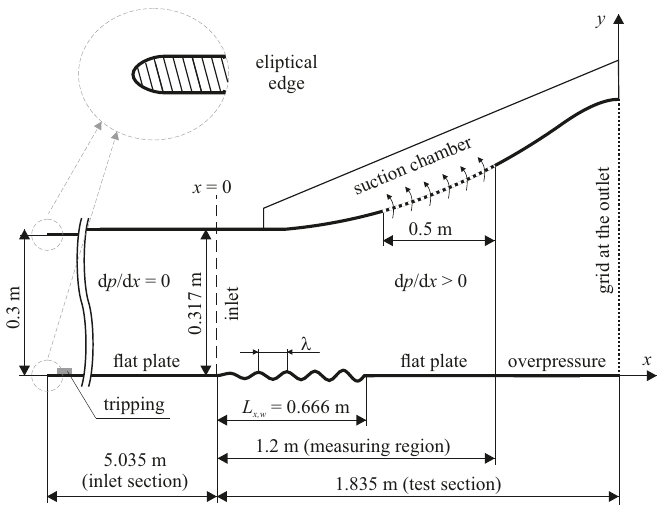}
    \caption{Schematic of the experimental stand from \cite{kaminski2024numerical}. }
    \label{fig:exp}
\end{figure}

\begin{figure}[h!] 
    \centering
    \includegraphics[angle=90, width=0.75\linewidth]{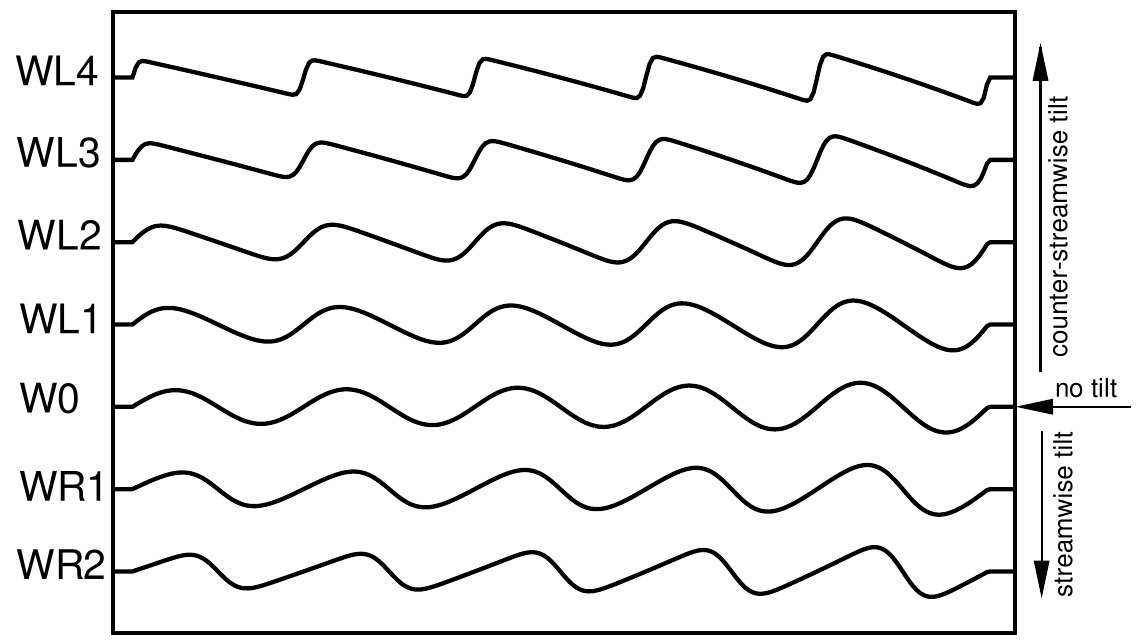}
    \caption{Waviness shapes. }
    \label{fig:shapes}
\end{figure}

Figure~\ref{fig:exp} shows a schematic of the experimental setup employed by Dróżdż et al.~\cite{drozdz2021effective} and later investigated by Kamiński et al.~\cite{kaminski2024numerical} using LES. In this setup, the flow develops along a 5035 mm long flat plate, which ensures that before the APG section, i.e. at $x=0$ (see Fig.~\ref{fig:exp}), TBL is well-formed. The section with the 2D waviness ($y=A\sin(x 2\pi N_\lambda/L_{x,w})$, $N_\lambda=5$ - number of periods) has the total length $L_{x,w}$ of 0.666 m ($\lambda = 0.1332$ m - the waviness period length). The amplitude $A$ changes along the waviness and grows according to the formula $A(x)=0.00366x^2 + 0.000614x + 0.003351$~m. For $Re_{\tau}=4000$ considered in ~\cite{drozdz2021effective}, this wall shape ensured a constant $A(x)/\delta(x)$ and led to the maximum increase in  $\tau_{w}$ ($
\Delta\tau_{w}=32$~\% compared to a flat wall). However, as shown in \cite{kaminski2024estimation} it turned out to be not optimal for $Re_{\tau}=2500$ with a larger $\delta(x)$. Parametric studies performed with a scaling factor $A_0$ ($A(x)=A_0\cdot(0.00366x^2 + 0.000614x + 0.003351)$~m) showed that with $A_0=1.2$,  a $20\%$ higher $\tau_{w}$ is obtained compared to the basic wall waviness ($A_0=1.0$). In this research, we maintain $A_0=1.2$ and demonstrate that significantly higher $\tau_w$ values can be achieved by tilting the waviness. The results suggest that future research for different $Re_{\tau}$ should treat the wavy wall shape and its amplitude equally important. To obtain a tilted waviness shape, an iterative procedure is proposed according to the following formulas: 

\begin{eqnarray}\label{eq:radii}
    y_0&=&\sin \left(x\frac{2\pi N_\lambda}{L_{x,w}} \right); \quad \textrm{initial sinusoidal waviness} \\
    y_{i}&=&\sin \left(x\frac{2\pi N_\lambda}{L_{x,w}} + \frac{y_{i-1}}{s}\right) + {y_{i-1}}\quad i=1,2,...,I\\
    y_{final}&=&A\times \left(\frac{y_I}{\max(|y_I|)}\right);\label{eq:radii-3}\quad \textrm{normalization, scaling}
\end{eqnarray}
where $s$ is the parameter defining the level of tilt and $I$ denotes the number of iterations $(i=1,2,...,I)$. An increase in $I$ causes the waviness to become less smooth. Shapes of the wavy walls considered in this study are shown in Fig.~\ref{fig:shapes} while Table~\ref{Tab:parameters} outlines the parameters used in the iterative procedure (\ref{eq:radii}-\ref{eq:radii-3}). 

The basic configuration is denoted as W0. The cases WL1-4 and WR1-2 represent tilted WW configurations with counter-streamwise and streamwise tilting of waviness crests, respectively. Please note that the flat channel configuration, marked in Table~\ref{Tab:parameters} as case F, is not included in Fig.~\ref{fig:shapes}. 

\begin{table}[ht]
\begin{tabular}{p{30pt}p{20pt}p{30pt}p{30pt}p{30pt}p{30pt}p{30pt}p{30pt}p{30pt}}
\hline
\hline
Case & F & WL4 & WL3 & WL2 & WL1 & W0 & WR1 & WR2\\ \hline
$s$ & - & 1 & 1 & 1 & 2 & - & -2 & -1\\
$I$ & - & 4 & 3 & 2 & 2 & 0 & 2 & 2\\
\hline\hline
\end{tabular}
\caption{Parameters of the iterative procedure (\ref{eq:radii}) for WW configurations.}
\label{Tab:parameters}
\end{table}

\section{Numerical model and boundary conditions \label{methods}}

The simulations are conducted using the ANSYS Fluent software, employing the Wall-Adapting Local Eddy-viscosity (WALE) subgrid model. The spatial discretization of the Navier-Stokes equations was accomplished with a bounded 2$^{\text{nd}}$ order central differencing scheme. A 2$^{\text{nd}}$ order scheme is also used for the discretization of the pressure gradient. Pressure-velocity coupling is enforced through the SIMPLE algorithm and 2$^{\text{nd}}$ order implicit time integration method is applied.  

\begin{figure}[h!] 
    \centering
    \includegraphics[width=0.95\linewidth]{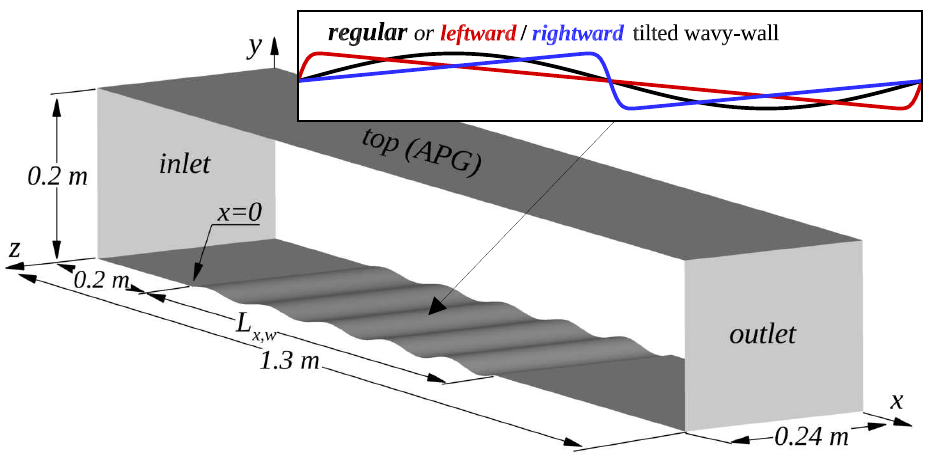}
    \caption{Computational domain.}
    \label{fig:domain}
\end{figure}

The computational domain, shown in Figure~\ref{fig:domain}, is smaller than the geometry of the experimental stand and it represents the measuring region (see Fig.~\ref{fig:exp}) of the length $L=1.3$ m. Such a simplification is necessary to reduce the computational cost of the simulations. With the present setting, the time required for modelling a single test case on a 96CPUs cluster equals 20 days. It covers an initial transient phase when the flow develops and the time-averaging period lasting over $3.6$~s that corresponds to $40$ flow-through times defined as $L/U_{x,\infty}$, where $U_{x,\infty}=14.42$~m/s is the free-stream velocity.

\begin{figure}[h!] 
    \centering
    \includegraphics[width=0.9\linewidth]{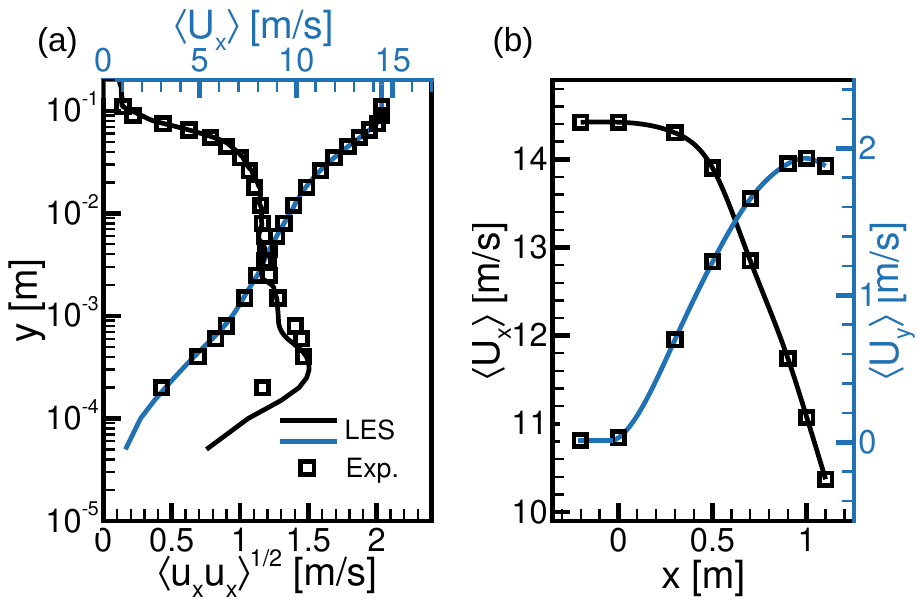}
    \caption{The approximated profiles: for the inlet boundary (a) and for the top boundary (b) generating the APG condition.}
    \label{fig:inlet}
\end{figure}

The air flows from the inlet at $x=-0.2$ and leaves through the outlet located at $x=1.1$ m.
Figure~\ref{fig:inlet}a shows the profiles of the mean streamwise velocity and its fluctuation imposed on the inlet. On the top boundary, the velocity fluctuations were negligibly small and therefore only the mean streamwise $\langle U_x\rangle$ and wall-normal $\langle U_y\rangle$ velocity components were assumed as the boundary conditions (Fig.~\ref{fig:inlet}b). The side boundaries, spaced by $z=0.24$ m, were assumed periodic and the bottom wall was treated as non-slip. At the outlet, the atmospheric pressure ($101325$ Pa) was assumed. 

The above-outlined solution procedure and the correctness of using simplified geometry has been thoroughly validated in Ref.~\cite{kaminski2024numerical} and will not be repeated here. The performed analysis included tests of the influence of a mesh density and time-step on the obtained solutions. The mean velocity and fluctuation profiles extracted at various channel locations agreed very well with the measurements. 
It has been shown that a multi-layer block-structured mesh consisting of $41.1\times10^{6}$ cells ensures virtually grid-independent results. In the presently analysed cases, the meshes slightly differ depending on the waviness shape but they all consist of $41.1\times10^{6}$ cells. The complexity of these configurations is similar to the complexity of the problems studied previously, and it is expected that the simulations reported in this paper are as accurate as those reported earlier.

\begin{figure}[h!] 
    \centering
    \includegraphics[width=0.85\linewidth]{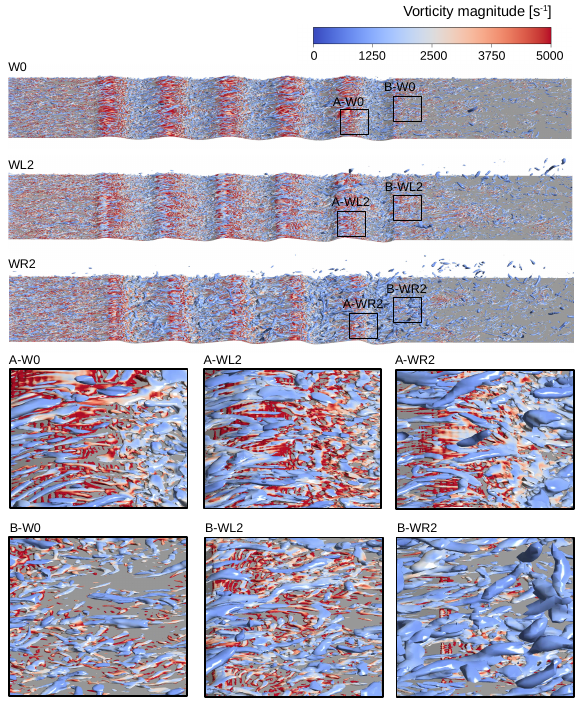}
    \caption{Isosurface of the Q-parameter, coloured by the vorticity magnitude, in the near wall region in the cases W0, WL2 and WR2.}
    \label{fig:qparam}
\end{figure}

\section{Results \label{results}}

As discussed in the introduction, the main goal of the research is to increase $\tau_w$, which will ensure the delay or complete elimination of TBL separation under the influence of a strong positive pressure gradient. On the other hand, it is known that any interference with the shape of the streamlined surface can contribute to the total drag. The purpose of this paper is to examine the effect of WW tilt on $\tau_w$ and the drag coefficient $C_d$ while maintaining the previously defined amplitude.  

First, we discuss general flow dynamics and identify differences between the solutions in selected cases based on the instantaneous results. Then we focus on the differences in $\tau_w$ and $C_d$, and try to explain them based on velocity and pressure distributions. Figure~\ref{fig:qparam} shows a near-wall flow region with isosurfaces of the Q--parameter colored by the vorticity modulus obtained for the cases W0, WL2 and WR2. The Q--parameter is defined as $Q = 0.5(\Omega_{ij} \Omega_{ij} - S_{ij} S_{ij})$, where $\Omega_{ij}$ and $S_{ij}$ are antisymmetric and symmetric parts of the velocity gradient tensor. The subfigures present an enlarged view of the flow near the tops of the last waviness period and the beginning of the flat wall section. It can be seen that large values of the vorticity occur mainly on the tops of the waviness. This is due to a strong velocity gradient close to the wall being the effect of a flow acceleration in this region. The Q--parameter is often used as an indicator of vortical structures and here it shows the simultaneous occurrence of tiny vortices and larger longitudinal structures, which break up on the downhill walls. On the waviness tops, the differences between the particular cases are small, with dominance of the longitudinal structures noted at the top of the wave. Conversely, at the beginning of the flat wall sections, the differences are pronounced. Although the flow manifests strongly turbulent behaviour in all cases, the sizes of vortices significantly change in particular configurations. The largest vortical structures occur in the case WR2 while for WL2, a much higher proportion of smaller turbulent scales and elevated levels of vorticity are noted.

\begin{figure}[ht] 
    \centering
    \includegraphics[width=0.85\linewidth, trim={0 2 0 0},clip]{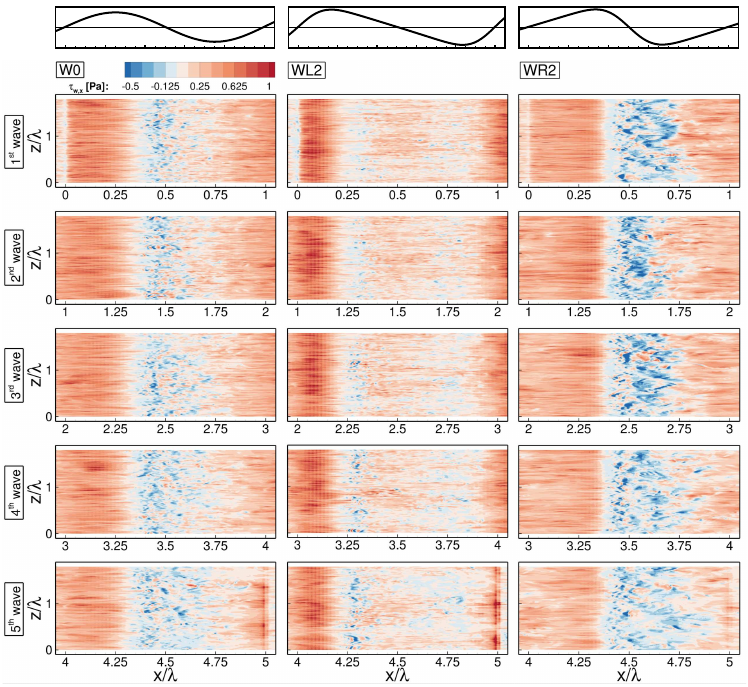}
    \caption{Contours of $\tau_{w,x}$ on the wall  in the cases W0, WL2 and WR2. The `x-z' coordinates are normalized by the waviness length $\lambda$. The upper schematics plots show the wall shapes.}
    \label{fig:tauw_cont}
\end{figure}

\begin{figure}[h!] 
    \centering
    \includegraphics[width=0.95\linewidth, trim={0 0 0 0},clip]{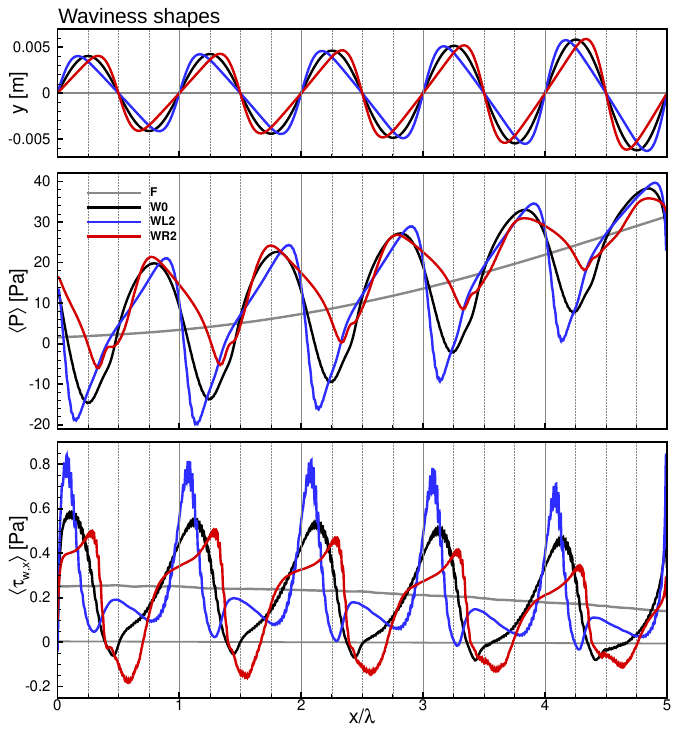}
    \caption{Profiles of $\langle\tau_{w,x}\rangle$ and $\langle P\rangle$ along the wavy wall section in the cases W0, WL2 and WR2. The lower figure shows the wall shapes in the waviness section.}
    \label{fig:tauw_profiles}
\end{figure}

\begin{figure}[h!] 
    \centering
    \includegraphics[width=0.75\linewidth, trim={0 0 0 0},clip]{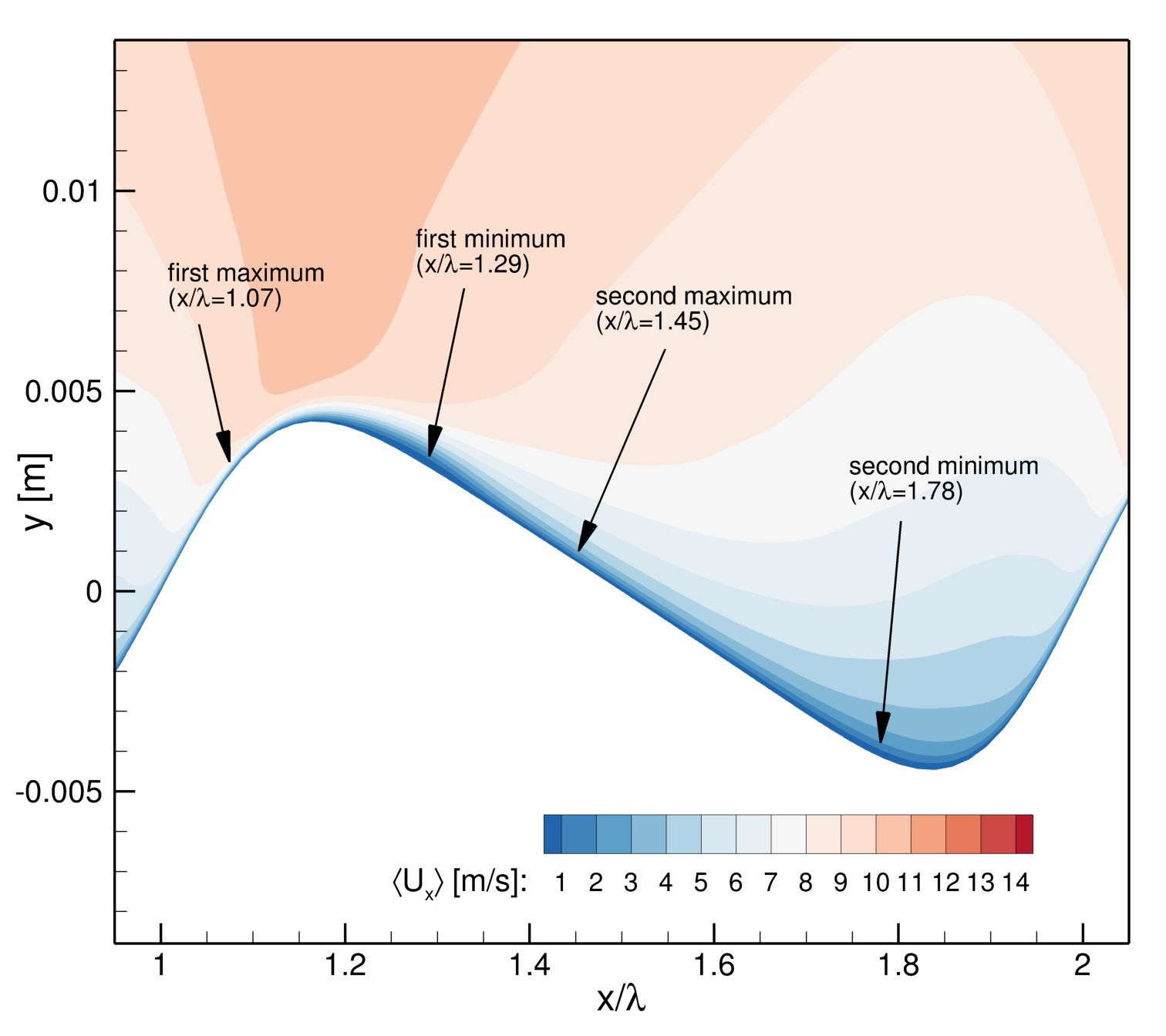}
    \caption{Contour maps of $\langle U_x \rangle$ for the case WL2, depicting near-wall area of the 2$\textsuperscript{nd}$ waviness period.}
    \label{fig:cont-n2}
\end{figure}

Figure~\ref{fig:tauw_cont} shows how the observed differences in the Q--parameter translate to the instantaneous streamwise component of the wall shear stress $\tau_{w,x}$ distribution in subsequent waviness periods. Comparing the cases W0 and WR2, although the latter exhibits larger negative $\tau_{w,x}$ values ($|\tau_{w,x,W0}|<|\tau_{w,x,WR2}|$), these solutions are qualitatively similar. The regions with $\tau_{w,x}<0$ occur on the inclined wall parts and starting from the 2$\textsuperscript{nd}$ waviness period their streamwise extents grow in the subsequent waves. Note, that when $\tau_{w,x}<0$ the instantaneous flow separation occurs. 
In the case of WL2, unlike in the cases W0 and WR2,  on the uphill slope, $\tau_{w,x}$ takes exceptionally large values. Downstream, the negative $\tau_{w,x}$ values are significantly smaller (($|\tau_{w,x,WL2}|<|\tau_{w,x,W0}|<|\tau_{w,x,WR2}|$)), and although they occur almost over the entire downhill length, it should be noted that the localisations of regions with $\tau_{w,x}<0$ are less frequent. This is reflected in its time-averaged profiles ($\langle \tau_{w,x}\rangle$) shown in Fig.~\ref{fig:tauw_profiles}, where additionally the pressure distribution on the wall is presented. For the ease of interpreting the graphs, the top panel shows the shape and location of each wavy geometry. In the case with the flat wall (case F) $\langle \tau_{w,x}\rangle$ changes only slightly and decreases along the streamwise direction as the overall pressure increases. On the contrary, when the wall is wavy, both $\langle \tau_{w,x}\rangle$ and pressure exhibit strong variations along the waviness. In the WL2 configuration, the maxima of $\langle \tau_{w,x}\rangle$ are significantly higher than in the other cases as evidenced by the $\tau_w$ maps presented in Fig.~\ref{fig:tauw_cont}. This is related to a change in the local pressure, which strongly decreases on the uphill waviness sides, where the flow accelerates. Also note the variation of the wall shear stress on the downhill slope, where two local minima, for WL2, can be observed within each period (for example, at $x/\lambda=1.29$ and $x/\lambda=1.78$). As a result of the acceleration on the uphill slope and the strong change in wall curvature, the boundary layer moves away from the surface, generating first local minimum in  $\langle\tau_{w,x}\rangle$ and forming a small (temporary, as seen in Fig.~\ref{fig:tauw_cont}) separation bubbles. The second local minimum in $\langle\tau_{w,x}\rangle$ is formed just before reaching the lowest part of the wave cavity, slightly preceding the local pressure maximum that is seen for the same case (WL2). This phenomenon is well visualised in Fig.~\ref{fig:cont-n2} depicting the mean velocity field over the 2$\textsuperscript{nd}$ wave of WL2. 

\begin{figure}[h!] 
    \centering
    \includegraphics[width=0.95\linewidth, trim={0 0 0 0},clip]{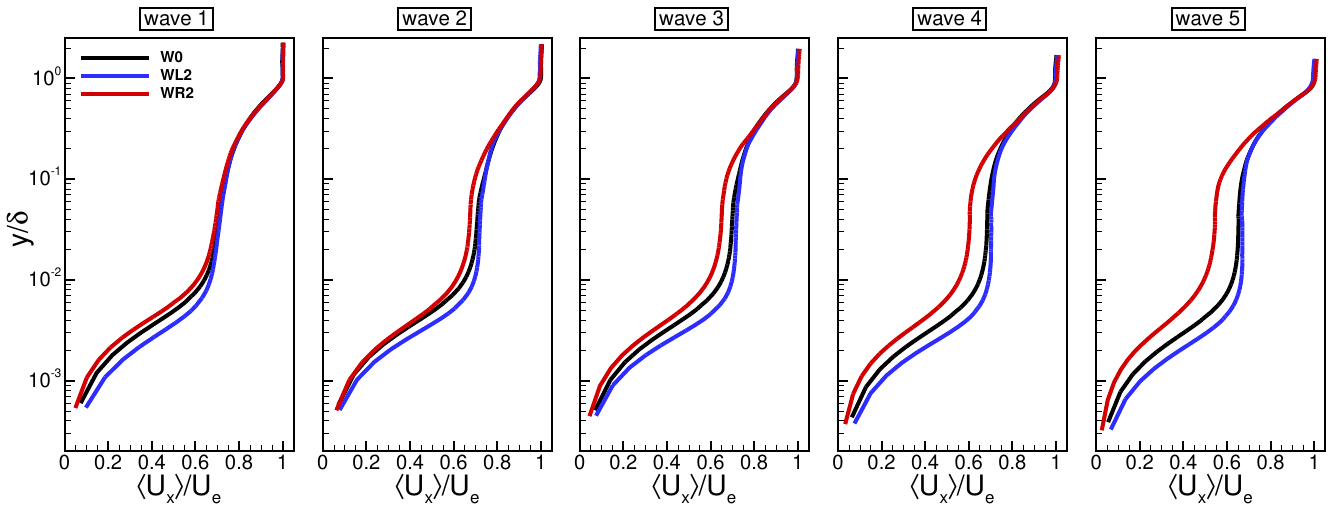}
    \includegraphics[width=0.95\linewidth, trim={0 0 0 0},clip]{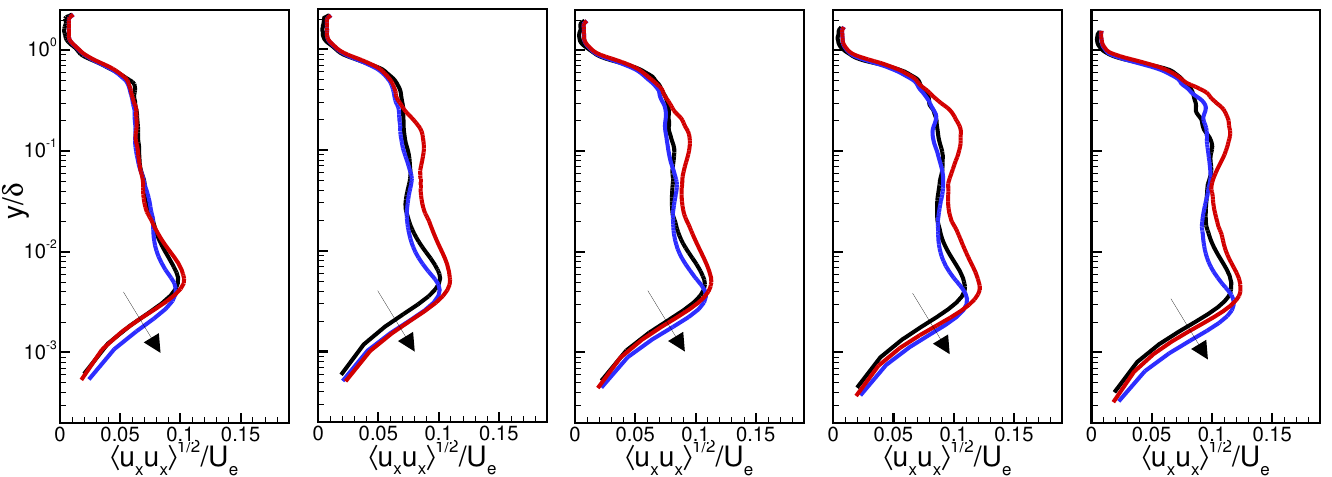}
    \caption{Profiles of $\langle U_x\rangle/U_e$ and $\langle u_xu_x\rangle^{1/2}/U_e$ along the vertical direction on the tops of the waviness in the cases W0, WL2, and WR2. }
    \label{fig:U_profiles}
\end{figure}

Figure~\ref{fig:U_profiles} shows the profiles of the streamwise velocity component $\langle U_x \rangle/U_e$ along the wall-normal direction on the tops of the waviness. The presented results are normalized by the boundary layer thickness ($\delta$), taken as the distance from the wall to the point where the velocity $U_e$ equals 99\% of $U_{x,\infty}$ at a particular streamwise $x-$location. It can be seen that in the case WL2 the velocity near the wall takes the largest values among the presented solutions and the differences increase in the subsequent waviness. In addition, these differences are also seen further away from the wall for successive periods, which is especially evident between WR2 and the other two cases. Note here the unfavourably strong deceleration of the flow in the inner zone of TBL for the case WR2 which had a direct attenuating effect on the wall shear stress. The acceleration of the flow on the tops causes the rise of the wall-normal velocity gradient resulting in the observed maxima of $\langle \tau_{w,x}\rangle$. Their values are sufficient to prevent the local flow separation ($\langle \tau_{w,x}\rangle<0$) on smoothly inclined walls of the downhill slopes. The velocity fluctuation distributions shown in the bottom row of Fig.~\ref{fig:U_profiles} indicate an increased rate of momentum transfer normal to the wall (see arrows in Fig.~\ref{fig:U_profiles}), explaining the increase in velocity in the near-wall region and the rise of $\langle \tau_{w,x} \rangle$ for the WL2 case. In this regard, the worst solution is found for the configuration WR2, where the regions with $\langle \tau_{w,x}\rangle<0$ are wide. The reason for that seems to be a too small pressure drop on the uphill waviness part, which compared to the cases W0 and WL2 is the least inclined. Simultaneously, the flow accelerates less than in the W0 and WL2 cases (see Fig.~\ref{fig:U_profiles}) and near the tops, the maxima of $\langle \tau_{w,x}\rangle$ do not exceed even those obtained in the case W0. In effect, the flow separates on the abrupt downhill waviness parts.

\begin{figure}[ht] 
    \centering
    \includegraphics[width=0.75\linewidth]{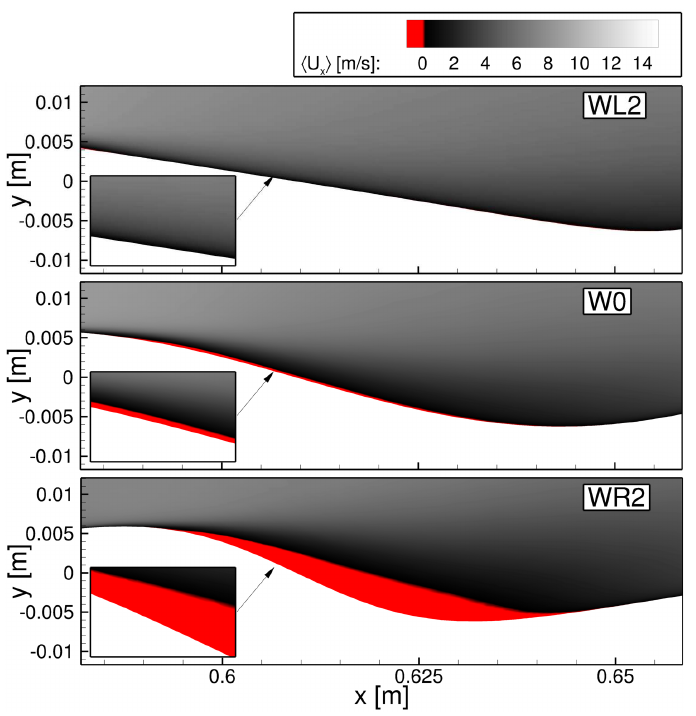}
    \caption{Contour maps of $\langle U_x \rangle$ at the last waviness trough.}
    \label{fig:separation}
\end{figure}
 
The separation areas, determined based on $\langle \tau_{w,x}\rangle < 0$, and previously discussed (see Figs.~\ref{fig:tauw_cont}~and~\ref{fig:tauw_profiles}), are illustrated in Fig.~\ref{fig:separation} as red zones (where $\langle U_x \rangle < 0$) on the streamwise velocity contour maps. These are depicted specifically for the 5$\textsuperscript{th}$ wave to emphasize the magnitude of detachment. Consistent with the observations from Figs.~\ref{fig:tauw_cont}~and~\ref{fig:tauw_profiles}, the largest separation zone appears for WR2, followed by W0. Additionally, although $\langle \tau_{w,x}\rangle < 0$ is observed for WL2 around $x/\lambda = 4.27$ (see Fig.~\ref{fig:tauw_profiles}), it is too minor to be visible on the $\langle U_x \rangle$ contours. This lack of detachment for WL2 is believed to lead to its lowest $C_d$ and possibly contributes to a significant improvement of $\tau_{w,x}$ enhancement over the sinusoidal waviness W0. This aspect is discussed in more detail later.

The above-discussed differences between the instantaneous results and between the profiles of the time-averaged pressure, streamwise velocity and $\tau_w$ along the waviness affect the flow behaviour downstream in the APG region. To minimize the risk of the TBL separation or postpone its occurrence, the $\tau_w$ values should be possibly large. In this respect, as an indicator of the performance of the WW geometries, we use $\Delta\langle\tau_{w,x}\rangle$ or $\Delta\langle\tau_{w,x}\rangle_{avg}=\overline{(\langle\tau_{w,x}\rangle_{WW}-\langle\tau_{w,x}\rangle_{F})/\langle\tau_{w,x}\rangle_{F}}$. Here, the triangular brackets represent the time averaging operator, the overline is the operator of the spatial averaging over the streamwise distance $0.7$ m $<x<1.1$ m, while $\langle\tau_{w,x}\rangle_{WW}$ and $\langle\tau_{w,x}\rangle_{F}$ indicate the wall shear stress computed for wavy and flat walls, respectively.

\begin{figure}[ht] 
    \centering
    \includegraphics[width=0.7\linewidth]{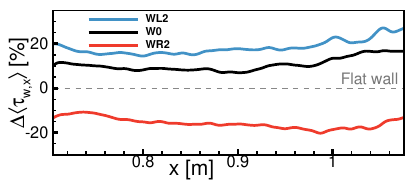}
    \caption{Profiles of: $\Delta\langle \tau_{w,x} \rangle$ for W0, WL2 and WR2 (a); $C_d$, $\Delta\langle\tau_{w,x}\rangle_{avg}$ (b) and $C_p$ with $C_f$ for all considered WW configurations.}
    \label{fig:dtw}
\end{figure}

Figure~\ref{fig:dtw} shows local values of $\Delta\tau_{w}$ for cases WL2, W0 and WR2. It can be seen that $\Delta\langle\tau_{w,x}\rangle$ levels for each configuration stay relatively constant but their values significantly differ from each other not only in the absolute levels but also in the sign. For WR2, $\Delta\tau_{w}$ is negative, which means that the overall value of $\tau_{w}$ is reduced and so the effect is detrimental compared to the case F. The benefits of using WW are seen for W0 and WL2 for which a growth in $\tau_{w}$ is exceptionally high. 

\begin{figure}[ht] 
    \centering
    \includegraphics[width=0.7\linewidth]{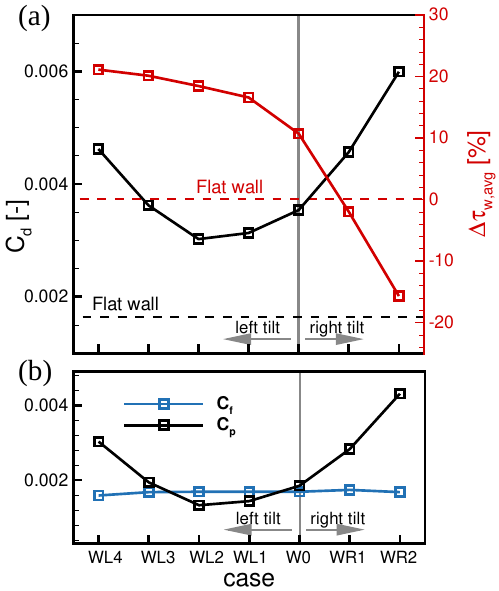}
    \caption{Profiles of: $C_d$, $\Delta \tau_{w,avg}$ (a) and $C_p$ with $C_f$ (b) for all considered WW configurations.}
    \label{fig:tauw-cd}
\end{figure}

Figure~\ref{fig:tauw-cd}a shows $\Delta\langle\tau_{w,x}\rangle_{avg}$ for all considered configurations (red line with symbols). Moreover, the presented results are supplemented with the corresponding values of $C_d$, computed for the waviness section, i.e., $0$~m $<x<0.666$~m. Note, that $C_d$ is the crucial parameter from the point of view of energy savings, and the smaller it is the better. In Fig.~\ref{fig:tauw-cd}b, the total drag coefficient is divided into its viscous and pressure contributions ($C_f$, $C_p$). The latter, usually called the shape drag coefficient, allows assessing how the shape of the body (here, the waviness) affects the total drag force.

For convenience, red and black dash lines representing $\Delta\langle\tau_{w,x}\rangle_{avg}=0$ and $C_d$ obtained for the case F are also included in Fig.~\ref{fig:tauw-cd}b to serve as a reference.
As can be seen from Fig.~\ref{fig:tauw-cd}a, the basic sinusoidal waviness geometry W0 ensures $\Delta\langle\tau_{w,x}\rangle_{avg}=10.7\%$, which is already a significant achievement. However, this can be substantially improved by tilting the waviness. When looking at the results obtained for the tilted WWs, it can be seen that the configurations with leftward tilt (WL1-WL4) cause an increase in $\Delta\langle\tau_{w,x}\rangle_{avg}$, while the cases with rightward tilt (WR1, WR2) lead to a drop in $\Delta\langle\tau_{w,x}\rangle_{avg}$. Worth noticing is also a considerable gradual growth in $\Delta\langle\tau_{w,x}\rangle_{avg}$ with increasing counter-streamwise tilting of the waviness WL1 $\to$ WL4. The changes in $\tau_{w,avg}$ are accompanied by the alteration in $C_d$. It is commonly known that, compared to the flat wall, the waviness causes an increase in $C_d$~\cite{cherukat1998direct, sun2018direct, kuhn2010large, tyson2013numerical, kruse2006structure, hamed2015turbulent, elsner2022experimental, akselsen2020langmuir, de1997direct, koyama2007turbulence, yoon2009effect, fujii2011turbulence, ghebali2017turbulent, segunda2018experimental, hamed2017turbulent}, however, this behavior can be controlled to some extent, as demonstrated in the present work. The distribution of $C_d$ exhibits a local minimum for the case WL2 which is approximately 14\% lower than in the case W0. Moreover, WL2, directly followed by WL1, display the dominance of viscous over pressure drag component (i.e. $C_f>C_p$, see Fig.~\ref{fig:tauw-cd}b), which is particularly interesting as it indicates that both of these WWs approach the shape of a streamlined body. For WL2 $\Delta\langle\tau_{w,x}\rangle_{avg} = 18.4\%$, which translates to 72.4\% higher $\tau_w$ than obtained for case W0. Assuming $\Delta\langle\tau_{w,x}\rangle_{avg}$ as the main indicator of the wavy wall performance, one can also see that it can be increased even further than in configuration WL2. In cases WL3 and WL4 $\Delta\langle\tau_{w,x}\rangle_{avg} = 20.11\%$ and $\Delta\langle\tau_{w,x}\rangle_{avg} = 21.57\%$, which compared the case W0 correspond to 88.1\% and 101.7\% rise in $\tau_{w,avg}$. This, however, is connected with an unfavourable increase in $C_d$ in which the shape drag dominates (see Fig.~\ref{fig:tauw-cd}b). The reason for the increase in $C_d$ and, in fact, its component $C_p$ for rightward tilt configurations is the presence of a large area of separation, as demonstrated in Fig.~\ref{fig:separation}. Nevertheless, the WL3 configuration is still better than W0, as $C_d$ in these two cases are almost identical. 
In the configuration WL4, $C_d$ rises significantly, and assessing the superiority of WL4 over the W0 case depends on the actual needs. If the increase in $\Delta\langle\tau_{w,x}\rangle_{avg}$ is the priority, the configuration WL4 proves to be the best among all the cases analysed.

\begin{figure}[ht] 
    \centering
    \includegraphics[width=0.95\linewidth]{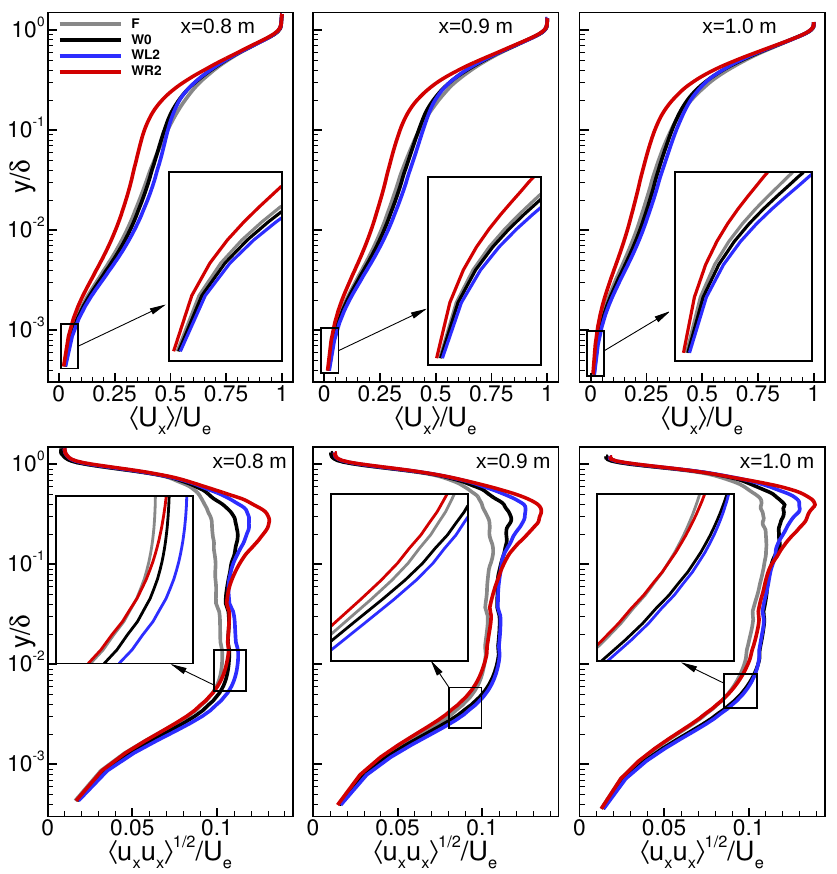}
    \caption{Profiles of $\langle U_x \rangle/U_e$ and $\langle u_xu_x \rangle^{1/2}/U_e$ at $x=0.8$ m, $x=0.9$ m, $x=1.0$ m.}
    \label{fig:profiles}
\end{figure}

In order to better explain the differences in $\tau_w$ levels, we analyze the mean velocity and its fluctuation distributions on the flate plate region in the measuring section as shown in Fig.~\ref{fig:exp}. 
Figure~\ref{fig:profiles} presents profiles of the streamwise velocity component $\langle U_x \rangle/U_e$ (upper row) and its fluctuation $\langle u_x u_x \rangle^{1/2}/U_e$ (bottom row) along the wall-normal direction for three different streamwise locations ($x=0.8$ m, $x=0.9$ m, $x=1.0$ m, which correspond to $0.124$ m, $0.224$ m, and $0.324$ m downstream the WW section, respectively). 
As can be seen, except for the WR2 configuration, the mean velocity profiles are roughly very similar. The pronounced velocity deficit for WR2 and thus the low velocity gradient at the wall translate into a low $\tau_{w,x}$. It should be noted that among the other three cases, WL2 exhibits the highest velocity gradient at the wall, which implies enhanced wall shear stress. These results are consistent with changes in velocity fluctuation profiles. It is apparent that fluctuations in the near-wall region ($y/\delta \le 0.01$) are the highest for WL2 which is accompanied by an increase in vorticity near the wall, as shown in Figure~\ref{fig:qparam}. This fluctuation increment, according to~\cite{hutchins2007large, RICCO2021100713}, is the result of the downwash of a high momentum fluid toward the wall by large vortical structures active in the outer layer. In contrast, $\langle u_xu_x \rangle^{1/2}$ for the F and WR2 cases is low in this region despite the fact that the level of $\langle u_xu_x \rangle^{1/2}$, i.e., the so-called outer peak, is highest in the outer layer. It is clear, however, that this does not translate into momentum transfer to the wall. All already described effects can be consistently noted at all considered downstream locations, i.e., $x$-distances. 

To sum it up, this paper has shown that by employing a waviness with a suitably adjusted tilt, a considerable enhancement in the wall shear stress, leading to the postponement of TBL separation, can be obtained with regard to a waviness without tilt. Moreover, this improvement is accompanied by a reduction of total drag. However, the benefits coming from tilting of the corrugation are not limited only to separation control, they can also be extended over, for instance, the heat transfer process. Namely, the considerable increase in fluctuations found in the outer region of TBL, provided by tilted waviness, could be beneficial for potential applications where heat transfer augmentation (by improving mixing) is sought after. For such applications, enhanced total drag is usually of lesser interest and so the corrugation with streamwise tilt (such as WR2) would be an excellent solution. Furthermore, if the goal is to enhance mixing while minimizing $C_d$, a wall with counter-streamwise tilt (like WL2) seems more attractive.

\section{Summary \label{conclusions}}
This paper demonstrates that counter-streamwise tilting of the wall waviness leads to a significant increase in wall shear stress compared to that observed with the classical sinusoidal wall shape. Moreover, this increase is accompanied by a local reduction in the drag force coefficient. At its minimum value, $C_d$, computed for the area occupied by the wavy wall, is 14\% smaller compared to the sinusoidal waviness, while the wall shear stress attains a level approximately 70\% higher. It turns out, this is not the limiting case. When a further enhancement of the wall shear stress is a priority to postpone separation, it can be achieved by steeper waviness tilting. This, however, comes at the cost of an increased drag coefficient. 

With the performed analysis of the flow in the wavy wall proximity, we can formulate the following advice. To increase the wall shear stress with a simultaneous decrease in the drag coefficient, one should shape the waviness in a way that prevents the occurrence of separation bubbles in the troughs. The waviness causes growth in the streamwise velocity fluctuations in the upper boundary layer region, making the flow more turbulent. This manifests in a larger near-wall streamwise velocity gradient, and thus, an increased wall shear stress.

On the other hand, configurations with streamwise waviness tilt, although disadvantageous from the viewpoint of flow separation postponement, provide a considerable increase in fluctuations in the outer region and thus an increase in turbulent kinetic energy. This makes them attractive for applications where enhanced mixing is beneficial, such as in heat transfer processes.

In real-life applications, such as wind turbine blades, where the wing does not operate in stable, constant conditions, i.e., the wind velocity may change rapidly, causing deviations in $Re$. Since passive flow control methods are susceptible to operating conditions, all aspects engaging their performance should be considered meticulously. This opens a space for future work aimed at further optimization of the corrugation to achieve the highest $\tau_w$ enhancement at minimal $C_d$. The results obtained in the present research are crucial for this task, as they highlight areas where the waviness can be further improved.

\section{Acknowledgements}

The investigation was supported by the National Science Centre under Grant No. UMO-2020/39/B/ST8/01449. The simulations were carried out using the PL-Grid computer infrastructure.

\newpage
\bibliographystyle{ieeetr}
\bibliography{bibliography}
%

\clearpage
\thispagestyle{empty}
\setcounter{page}{0}

\begin{singlespace}

\clearpage

\end{singlespace}

\end{document}